# FRAUDULENT ELECTRONIC TRANSACTION DETECTION USING DYNAMIC KDA MODEL

M.Vadoodparast[1], Prof. A. Razak Hamdan[1], Dr. Hafiz[1]

[1]Management Information System Dep, Faculty of Information Technology, University Kebangsaan Malaysia, 43600 Bangi, Selangor, Malaysia

*Abstract* – Clustering analysis and Datamining methodologies were applied to the problem of identifying illegal and fraud transactions. The researchers independently developed model and software using data provided by a bank and using Rapidminer modeling tool.

The research objectives are to propose dynamic model and mechanism to cover fraud detection system limitations. KDA model as proposed model can detect 68.75% of fraudulent transactions with online dynamic modeling and 81.25% in offline mode and the Fraud Detection System & Decision Support System. Software propose a good supporting procedure to detect fraudulent transaction dynamically.

## I. Introduction

Today's detecting and preventing fraudulent financial transactions especially in credit cards from huge volume of data are playing important role in the banking and financial institutions business. Many researches have used data mining algorithms to detect fraudulent transactions. Normally more than one million transactions are created daily, so detecting process in optimal way is a time consuming process and mostly is done offline in static operation, usually the batch processing is used in specific period like daily, weekly or monthly to discover the fraud. The second issue is the learning machine or supervised algorithm like classification relies on accurate identification of fraudulent and non-fraudulent transactions, however these information usually do not exists or limited. Also, it means preventing of happening fraudulent transaction do not occur in transaction time or the system using predefined rules and scenarios or static model. In order to fill this gap and needs of periodically update of rules to perform optimally, it is necessary to present dynamic models. Thus, the research objectives are to propose dynamic model and mechanism to cover these two issues. The standard data mining methodology is adopted in this research. Table 2 shows the researches have done by researcher based their country; we can see that United State has most part, based on Table 1, we show that the United State suffer for fruad problem with overally 42% in last years, it means US has good approch to manage this problem.

| Country | Cardholders Affected (Overall) | Cardholders Affected (Last 5 Years) |
|---|---|---|
| United States | 42% | 37% |
| Mexico | 44% | 37% |
| United Arab Emirates | 36% | 33% |
| United Kingdom | 34% | 31% |
| Brazil | 33% | 30% |
| Australia | 31% | 30% |
| China | 36% | 27% |
| India | 37% | 27% |
| Singapore | 26% | 23% |
| Italy | 24% | 22% |
| South Africa | 25% | 20% |
| Canada | 25% | 19% |
| France | 20% | 18% |
| Indonesia | 18% | 14% |
| Sweden | 12% | 11% |
| Germany | 13% | 10% |
| Netherlands | 12% | 8% |

Table 1. Cardholders Impacted by Fraud by Country [1]

| Country | Study | Method | Details |
|---|---|---|---|
| USA | Ghosh & Reilly(1994) | Neural networks | FDS (fraud detection system) |
| | Ezawa & Norton (1996) | Bayesian networks | Telecommunication industry |
| | Chan et al. (1999) | Algorithms | Suspect behavioral prediction |
| | Fan et al. (2001) | Decision tree | Inductive decision tree |
| | Maes et al. (2002) | Bayesian networks & neural networks | Credit card industry, back-propagation of error signals |
| UK | Bently et al. (2000) | Genetic programming | Logic rules and scoring process |
| | Wheeler & Aitken(2000) | Combining algorithms | Diagnostic algorithms; diagnostic resolution strategies; probabilistic curve algorithm; best match algorithm; negative selection algorithms; density selection algorithms and approaches |
| | Bolton & Hand (2002) | Clustering techniques | Peer group analysis and break point analysis |
| Germany | Aleskerov et al.(1997) | Neural networks | Card-watch |
| | Brause et al.(1999a) | Data mining techniques & neural networks | Data mining application combined probabilistic and neuro-adaptive approach |
| Canada | Leonard (1995) | Expert system | Rule-based Expert system for fraud detection (fraud modelling) |
| Spain | Dorronsoro et al.(1997) | Neural networks | Neural classifier |
| Korea | Kim & Kim (2002) | Neural classifier | Improving detection efficiency and focusing on bias of raining sample as in skewed distribution. To reduce "mis-etections". |
| Cyprus | Kokkinaki (1997) | Decision tree | Similarity tree based on decision tree logic |
| Singapore | Quah & Sriganesh (2007) | Neural networks | Self-Organizing Map (SOM) through real-time fraud detection system |
| Ukraine | Zaslavsky &Strizkak (2006) | Neural networks | SOM, algorithm for detection of fraudulent operations in payment system |

Table 2, summary of studies investigating different techniques in credit card fraud [2]

## II. Overall Process of Data Mining for fraud detection

The overall process of fraud detecting using data mining methods cotains following steps as showon in Figure 1.
- Gathering data of domain and related knowledge
- Selecting transactional dataset based on date and time or quantity or combination of both, customer based or customer group based.
- Preprocessing (Remove noise,Handling missing value,Transformation into suitable form for mining.
- Using Data mining technique which is usully searching patterns based on models such as classification,neural network or outlier recognition based on clastering technique.
- Pattern or outlier evaluation to identify representing knowledge
- Send the extraeted information to DSS in order to decide wither it normal or abnormal behavior

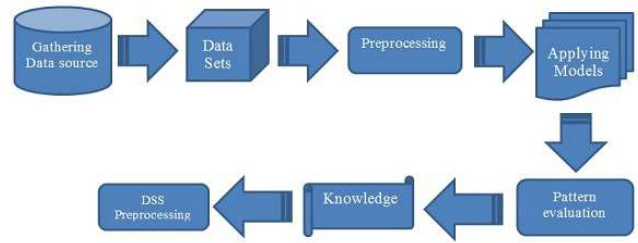

Figure 1. Overall process of fraud detection

Normally these process is done offline or staticly becuase the volume of transaction is huge and this process is time consuming, so providing a dynsmic model for huge volume of data is not esay and processing this model take time ,while the transaction done in less than mili second.

One of the most important challenge is, using supervised data mining technique like learning machine or classification relies on accurate identification of fraudulent and non-fraudulent transactions, however these information usually do not exists or limited or confidential . Financial institutions prefer to not disclose this kind of information and categorize them in high-risk data, so accessing to this kind of data is very restricted. Therefore the process has difficulty in step "Applying Models" and "Pattern evaluation", so the extracted knowledge might not be cover all fraud scenarios and it increase the error and decrease the accuracy and finally the Decision Support System (DSS) accuracy is decreased as well. So many researches have done to fill this gaps and present models or techniques to overcome these issues and enhance the DSS.

## III. Data Mining and clustering

In clustering problems usually, we have set of properties or dataset and looking for some similarity or dissimilarity based on some predefine criteria. This similarity criterion case by case is different for different problems. For example if the datasets are contiguous we can use Euclidean distance as similarity criteria[3], so every dataset will map in multidimensional space as point and each dimension represent one feature or property of dataset.

In clustering problems, there is no special class, actually, we do not have class factors as classifier and just based on similarity, the categorization and clustering will be done. The most similar records or dataset will group in same cluster, so the different clusters have less similarity to each other.

Because of we are not defining classifier for clustering algorithm and data do not labeled or tagged, this technique categorize as unsupervised techniques. The clustering results will analysis for extracting order or knowledge from clustered datasets. Clustering outputs reanalysis again in order to find discipline between clusters, the important point is that, always clustering work based on input properties or parameters, same dataset with different parameter might lead to different clustering results and it is not related directly to clustering algorithm.

The aim of clustering is minimize the Intercluster Distance and maximize the Intracluster Distance. (See figure 2 regarding this)

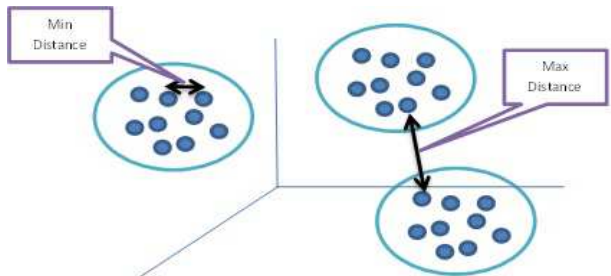

Figure 2. Clustering process Objectives
(adapted from Practical Datamining book,2012)

A good clustering method will produce high quality clusters in which:
• The intra-class (that is, intracluster) similarity is high.
• The inter-class (that is, intercluster) similarity is low.

The quality of a clustering result also depends on both the similarity measure used by the method and its implementation. The quality of clustering method measures, by its ability to discover some or all of the hidden patterns as well.

## IV. Schematic Overview of Clustering Process in this Paper

In Figure 3 the overall process and steps of fraud detection and DSS are presented.in first step, the historical repository database of previous customer transactions should be prepared and based on model required parameters, the preprocessing are applied. When new transaction comes, based on data window size (that is last 100 transactions, in this paper), customer dataset fetch from repository, and a new transaction is sent to the clustering model, in order to develop a customer behavior model. After applying model, the clustering results will send to a DSS in order to decide whether it suspicious transaction or the behavior is normal. The result evaluation is, based on genuine fraud cases as external dataset evaluation.

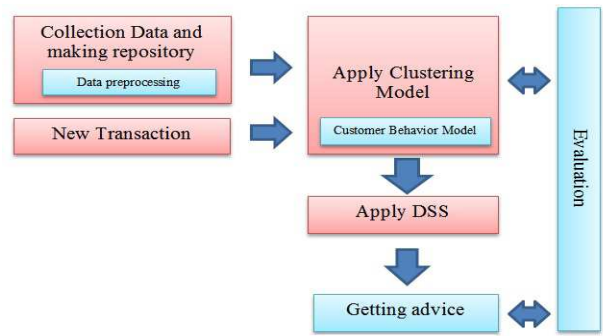

Figure 3. the overall process of FDS in this thesis

In this research, 3,609,618 real banking transaction data for 1015 customers were collected and 32 genuine fraudulent cases are used to compare and to evaluate the result.

The transaction data are preprocessed in order to improve the quality and process speed. The collected data contains 44 items per transaction and we used 8 items for modeling purpose. For the time accuracy the hour accuracy are considered, so to process the historical date so the transaction data should grouped, hour wised.

Data preprocessing step uses for optimizing quality of datasets, in clustering model. However, in this thesis the statistical data is used for all datasets without any elimination because we looking for outlier and abnormality in this model. If we remove outliers, we might lose suspicious transactions. After preprocessing, repository data is ready to use for customer behavior modeling.

In preprocessing data, we filtered the transaction data based on:
- The transaction should be from purchasing type group like retail transaction, bill payment or top up transaction
- The transaction should be settleted
- We extract :
    o  PAN for identify the customer
    o  TermID for identify the terminal id,normally the customers using same  place or same web payment in their transaction.
    o  MerchantID to identify the merchant ,normally customers using same merchant for their reguller shopping
    o  PosCondition to identify the paymnet device like POS , Mobile,Internet.normally customers have some habit in using media like mobile or POS.
    o  AffectiveAmount as transaction amount
    o  BusinessDate as transaction date
- We have processed BusinessDate and divided it in two fields: transaction Date and transaction Time based on transaction hour. Normally the customer make their transaction in similar date like end of month and usually in same hour ,especially for bill payment

The result of preprocessing dataset is shown in Table 3, that will be used in research data mining model as input repository.

| R | Filed Name | Type | Description |
|---|---|---|---|
| 1 | PrCode | Integer | Process code type of transaction |
| 2 | PAN | Varchar | Masked Card NO |
| 3 | TermId | Varchar | Terminal identifier |
| 4 | MerchantID | Varchar | Merchant identifier |
| 5 | PosCondtion | Integer | POS Operation type (Bill,Top up,…) |
| 6 | AffectiveAmount | Double | Affective transaction amount |
| 7 | TrxDate | Date | Transaction business date |
| 8 | TrxTime | Integer | Transaction Hour |

Table 3 dataset filed after preprocessing

All algorithms needs one tag as identifer to make it unieuqe in data set for each record so, we add one more lable as ID to identify each record throug and after processing.

## V.  KDA Clustering Model

As shown in Figure 4, the final proposed model as KDA clustering model is a combination of three clustering algorithm, K-MEANS, DBSCAN and AGGLOMERATIVE clustering algorithms that represented together as dynamic solution. When new transaction happened, the customer behavior model generate (including new transaction) and the customer dataset cluster with three clustering algorithm, K-MEANS, DBSCAN and AGGLOMERATIVE, it means each record will have three labels that will use to detect abnormality.

Each algorithm might use all or some parameters of prepared dataset. Suspicious transaction will be in the clusters with minimum members in K-MEANS, high LOF values in DBSACN and in a single node in AGGLOMERATIVE algorithm that appear and detected at least by two of clustering algorithms. It means if the new transaction detected by two or more algorithm in as suspicious transaction, it takes place in suspicious area and will potentially fraudulent transaction.

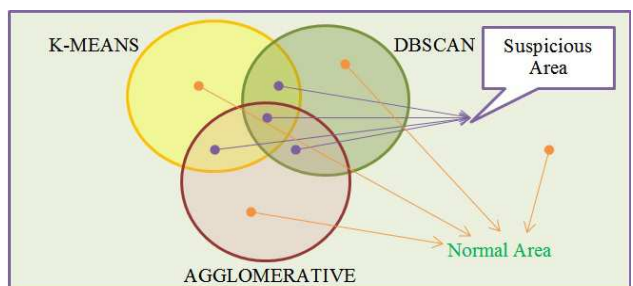

Figure 4. KDA Clustering Model diagram

In this model, the model processing happen parallel for each algorithm and the results will write to separated tables in database, so we can easily with comparing result detect abnormally in customer behavior.

K-MEANS good enough when, we able to define center points and define K as number of clusters and it can detect noise and outlier by measuring distance very good, we can find and optimize center point ( here named centroid) by repeating and rerunning the algorithm again on the result of previous execution. So, the problem of this algorithm is finding optimal K.

We can summarize K-MEANS steps as[3]:

- Input : K, number of cluster and n, objects dataset
- Output: set of K cluster with minimum squared errors criteria

Below are algorithm steps:
1) Pick a number (K) of cluster centers - centroids (at random)
2) Assign every item to its nearest cluster center (e.g. using Euclidean distance)
   - $d_E(x,y) = \sqrt{\sum_{k=1}^{n}(x_k - y_k)^2}$
   - $d_H(x,y) = \sum_{k=1}^{n}|x_k - y_k|$
   
   That, n is number of dimensions or number of dataset properties and $x_k$ and $y_k$ are k properties of $x$ and $y$ objects

3) Move each cluster centre to the mean of its assigned items
4) Repeat steps 2,3 until convergence (change in cluster assignments less than a threshold)

In the DBSACN algorithm, the number of the clusters not fixed or predefined, this algorithm looking for point with maximum density in their surrounding neighborhood and number of cluster specified dynamically. The one cluster based on density contains a set of objects that all Density-Connected to each other. That means any object outside these cluster consider outlier or noise.For detecting local outlier, a degree to each object will assign to be an outlier. This degree called the Local Outlier Factor (LOF) of an object. The degree depends on how the object is isolated with respect to the surrounding neighborhood Defining ε as surrounding neighborhood radius is very important in this algorithm because if assign small number, number of the clusters will increase and all data going to separated clusters and if assign big number all data going to one big The clusters will increase and all data going to separated cluster , so finding optimal ε is very important.

We can summarize DBSCAN steps as [3]:
1) Computing (k- distance of p)
2) Finding (k-distance neighborhood of p)
3) Computing(reachability distance, p wrt object o)
4) Computing (the local reachability density of p)
5) Calculation Local outlier factor of p

AGGLOMERATIVE algorithm works like tree, first consider each object as one cluster and then start combining these clusters together based on some criteria and make bigger cluster until all cluster combine and make big tree or meeting stop condition[4]. This algorithm works by comparing distance between all objects in same cluster together and divides the objects with maximum similarity in one cluster, and repeat processing with new cluster. In this algorithm if repeat cycle many times, all objects will be take place in one cluster separately and if we run it enough might the results not good to make decision regarding results. Taking place in one cluster is not a problem because the model represent tree and by analyzing tree we can take decision but it time consuming process and might run cycle hundred or more times, therefore stop condition is main issue of this algorithm.

| Algorithm | Main Issue |
|---|---|
| K-MEANS | Assigning proper K |
| DBSCAN | Defining proper $\varepsilon$ |
| AGGLOMERATIVE | Stop condition |

Table 4. Compare Clustering algorithms main issues

In proposed technique, the final decision make based on comparing of output of all algorithms together in order to decrease the errors and increase the accuracy K-MEANS is fast and the accuracy is good but it is static clustering, so we cover it with DBSCAN and AGGLOMERATIVE with dynamic cluttering. DBSACN is dynamic but if fraud happen out of ε radius cannot detect it, but K-MEANS and AGGLOMERATIVE able to detect noise in all distances. AGGLOMEARTIVE is dynamic but not enough fast and might put all object in one cluster specially when increase the parameters, but K-MEANS and DBSACN have stop condition. So, we can conclude these using algorithms together can cover each other to solve the fraud problem better.

## VI. Model Specification

For bulding customer behavioral model we select bellow items as K-MEANS dimensions, that means, this model has 6 dimensions.
- AffectiveAmount
- MerchantID
- PosCondtion
- PrCode
- TRXDate
- trxtime

Parameters:
- Number of attribute =6
- K=12
- Max runs=10
- Measure Type= Numerical Measure
- Numerical Measure = Euclidean Distance
- Max optimization step=100

in this model, it is set K=12 that is for last 3 months equal to 12 weeks, the purpose is cluster every week in one cluster if everything be normal, and n=100, that maximum number of transaction in last 3 months. In evaluation phase, we will evaluate the accuracy of K with Davies-Bouldin index calculation as performance evaluation for K-MEANS clustering and prove that the K=12 is the optimal.

For DBSCAN like K-MEANS, 6 items parameters are used and numerical measure as measure type, with Euclidean distance calculation to calculate dependency for detecting noise and outliers are used.
These 6 dimensions for this clustering include:

- AffectiveAmount
- MerchantID
- PosCondtion
- PrCode
- TRXDate
- trxtime

Parameters:
- Number of attribute =6
- Epsilon($\varepsilon$)=1000000
- Min points=1
- Measure Type=Numerical Measure
- Numerical Measure = Euclidean Distance

The minimum cluster object is set to 1, it means at least the output has one cluster, and $\varepsilon = 1000000$ that the minimum amount that important in banking system to inspect for fraud in fraud detection process (the currency is Rails and this amount equal to 100 Malaysian ringgits)

For AGGLOMERATIVE algorithm, we choose three dimensions for this clustering:
- AffectiveAmount
- TRXDate
- trxtime

Parameters:
- Number of attribute =3
- Mode=Average Link
- Measure Type=Numerical Measure
- Numerical Measure = Euclidean Distance

To simplify the complexity of this algorithm, the parameters are reduced to three fields, Average Link are recruited. Numerical measure as measure type with Euclidean distance is selected as well.

These three clustering algorithms works with numerical data not nominal, so converting data to numeric is one pre step before running the model. We used convertor adapter to convert nominal data to numeric, and all data converted to numeric.

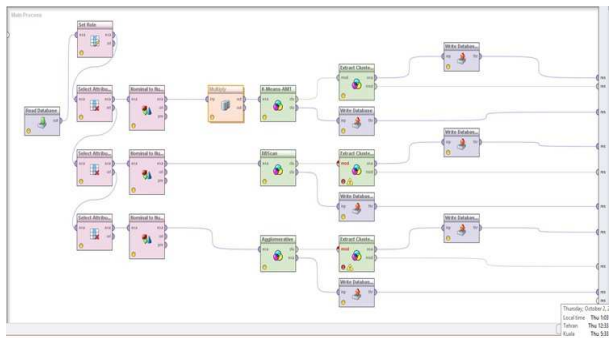

Figure 5. Implemented KDA model in RapidMiner

Figure 5 shows the implemented KDA dynamic model in RapidMiner software.

When new transaction happened, the customer behavior models generate for these three selected algorithms (including new transaction) and suspicious transaction will take place in shared space between at least two algorithms that usually, are in the clusters with minimum members and high LOF values or in a single node. So, the KDA model space is shared spaces between these three algorithms that each algorithm try to detect abnormality with different technique, on the other hand, overlapping areas are as desire area and required answer for fraud detection problem.

In this model, the model processing happened parallel for each algorithm, it means, we are checking distance, density and objects route link together in same time and then deciding regard occurred transaction, we try to see transaction from different perspective to make sure detecting process work optimally. The results of each algorithm write to separated tables in database, so we can easily detect abnormally in behavior with comparing result.

## VII. Discussing FDS&DSS Logic

Decision support system regarding fraud detection is a one of most important section in all financial organization, that wrong decision influence directly the business and it causes dissatisfaction in customer area. Therefore, the decision rules and policies are normally conservative and somehow managers prefer to inspect issues manually or just getting advices form Fraud Detection System (FDS) regarding stop online suspicious transaction specially when new scenarios happening. With growing fraudulent transactions in last years the approach of using automated FDS is increased and many FDS are developed.

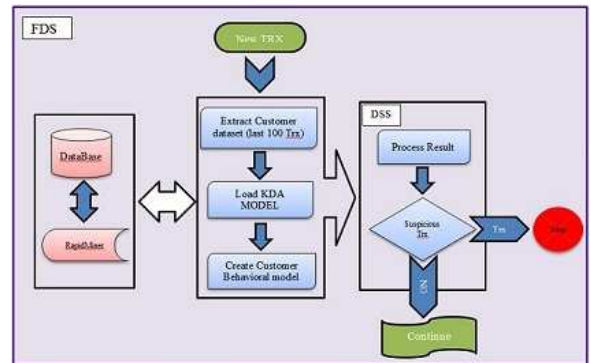

Figure 6. Proposed FDS diagram

In proposed DDS that use KDA model for detection fraud inspect the suspicious area and if transaction take place in this area the system will arise alert to advise user to inspect the transaction or stop it.

The DSS logic is simple and works as:
If trx(n) detected by K-MEANS(n) as Fraud then
   nK=1
  Else
nK=0
If trx(n) detected by DBSCAN(n) as Fraud then
   nD=1
Else
   nD=0
If trx(n) detected by AGGLOMERATIVE(n) as Fraud then
   nA=1
Else
   nA=0
If (nK and nD) or (nK and nA) or (nD and nA) then
   nF=1 ;
   SendAlert;
Else
   nF=0
   Continue;

Based on bank policy, the system can stop the suspicious transaction or just raise alert for user in order to inspect the transaction.

## VIII. Discussion & Results

As definition, we have defined:
- True Positive Rate (TPR) → Normal transactions and model detect normal
- False Positive Rate (FPR) → Abnormal transactions and model detect normal
- True Negetive Rate (TNR) → Normal transactions and model detect Abnormal
- False Negetive Rate (FNR) → Abnormal transactions and model detect Abnormal

This model aimed to increse True Positive Rate (TPR) and False Negetive Rate (FNR), it means increase accuracy regarding normal and abnormal transactions, and decrease False Positive Rate (FPR) and True Negetive Rate (TNR) means reducing errors. On the other hands, the system detect normal and abnormal transactions properly and reduce errors in this process.

we have run this model for 100 customer that already have normal transactions in Databases and investigate the results.

| R | Model | TPR | TNR |
|---|---|---|---|
| 1 | K-MEANS | 90 | 10 |
| 2 | DBSCAN | 84 | 16 |
| 3 | AGGLOMERATIVE | 88 | 12 |
| 4 | KDA Model | 96 | 4 |

Table 5 , Model results for Normal transactions

Results show , the KDA model can detect 96% of normal transaction properly. Logic of KDA is based on, if two model detect one transaction as normal transaction , the result will be normal and this optimazation is becuse of using more than one clustering technieque in the final model. We can see, if we using one clustering model, in best state the result will be 90% that related to K-MEANS, but here the KDA model accuracy is 96% , it means atleast 6% of normal transactions in K-MEANS detect as abnormal, on the other hand the KDA model optimize error of K-MEANS 6%, DBSCAN 12% and AGGLOMERATIVE 8% as well.

From other point of view, we can see, at least 6 transactions exist that K-MEANS algorithm cannot detect it properly but DBSCAN and AGGLOMERATIVE can detect them better. Figures 7-10 show model output from RapidMiner software.

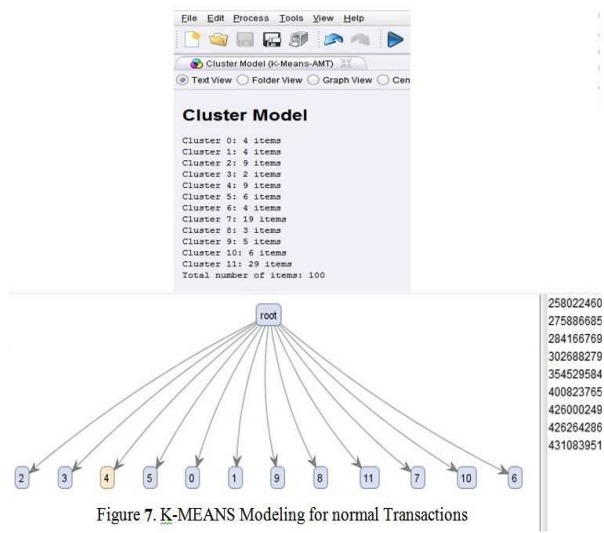
Figure 7. K-MEANS Modeling for normal Transactions

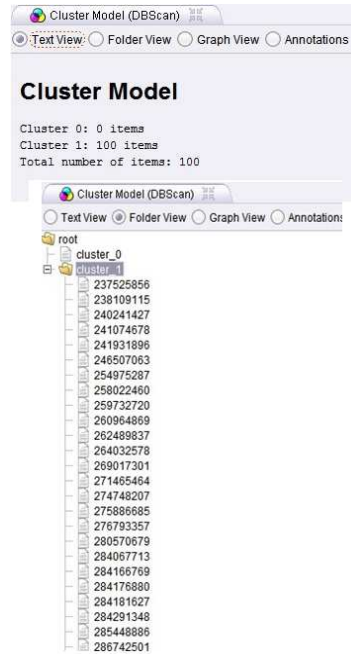
Figure 8. DBSCAN Modeling for normal Transactions

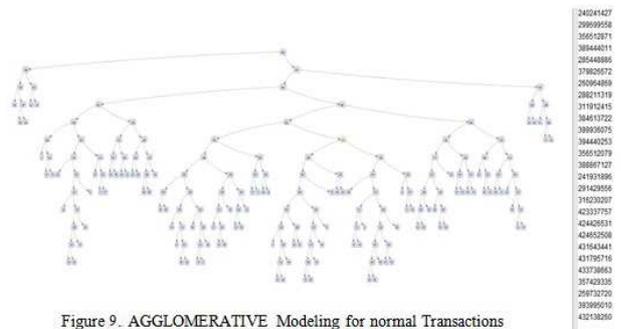
Figure 9. AGGLOMERATIVE Modeling for normal Transactions

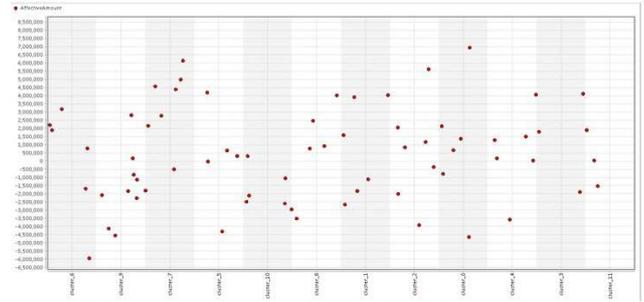
Figure 10. K-MEANS clustering for AffectiveAmount for normal Transactions

For testing the model with genuine cases as external evaluation ,we have run this model for 32 fraudulnet transaction,as mentioned previously ,the Database has 1015 customers information.

In first step , we have run the model for all historical customers data , in this period all fraudulent transaction is 16 and the model has detected 18 transaction as fraud , from this 18, 13 was correct , it means FNR=13 and TNR=5 and model could not detect 3 transations at all and detect them as normal transations, it means FPR=3.Table 5.3 shows the KDA model resultant is better then each model seperatly, we can see K-MEANS model is more sensetive thant two other models but the precision is lower (FPR, TNR is bigger) and agglomertive detection is less sensetive but false detection is better (FPR, TNR is smaller).Results are shown in Table 6.

| R | Model | Total Detect | TNR | FNR | FPR |
|---|---|---|---|---|---|
| 1 | K-MEANS | 21 | 10→62.5% | 11→68.75% | 5→31.25% |
| 2 | DBSCAN | 19 | 7→43.75% | 12→75% | 4→25% |
| 3 | AGGLOMERATIVE | 17 | 6→37.5% | 11→68.75% | 5→31.25% |
| 4 | KDA Model | 18 | 5→31.25 % | 13→81.25% | 3→18.75% |

Table 6.Model results for fraud detection

In next step , we test the model with genuine fraudulent cases in real time to see the result of dynamic modeling. We test the model with 16 frudulent transactions. The result are shown at Table 7

| R | Model | FNR | FPR |
|---|---|---|---|
| 1 | K-MEANS | 9 →56.25 % | 7→43.75 % |
| 2 | DBSCAN | 7 →43.75% | 9→56.25% |
| 3 | AGGLOMERATIVE | 8 →50% | 8→50% |
| 4 | KDA Model | 11→68.75% | 5→31.25 |

Table 7.Model results for real time fraud detection

The results show, the KDA model still have better results than each model separately, with combination of each model results with this logic: "if the transaction detected by two model as fraud so in KDA model consider as suspicious transaction", the final results is 68.75 %.of fraudlent transactions can be detected by this model.Figures 11-15 show model output from RapidMiner software.

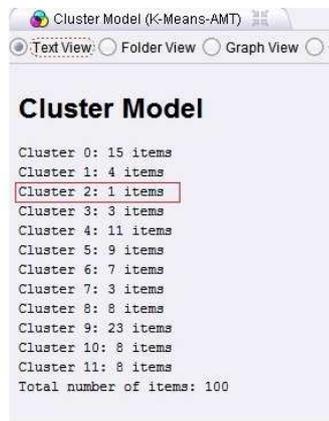

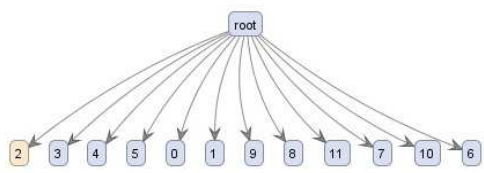

Figure 11. K-MEANS Modeling for abnormal Transactions

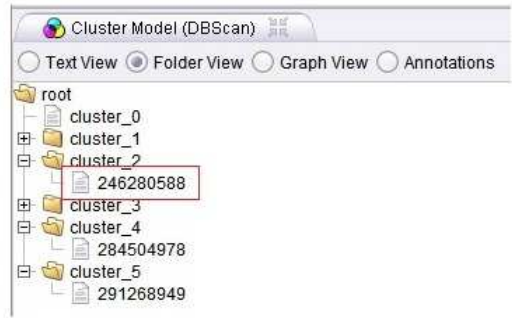

Figure 12 . DBSCAN Modeling for abnormal Transactions

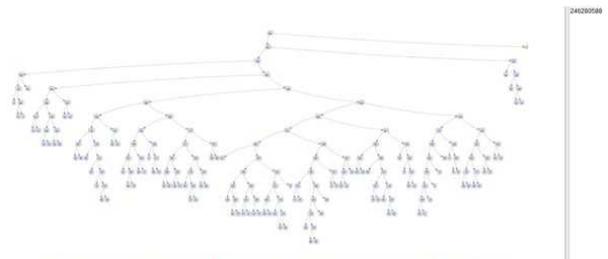

Figure 13 . AGGLOMERATIVE Modeling for abnormal Transactions

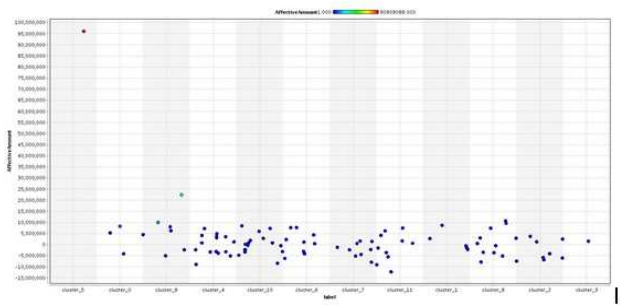

Figure 14. K-MEANS clustering for AffectiveAmount for abnormal Transactions

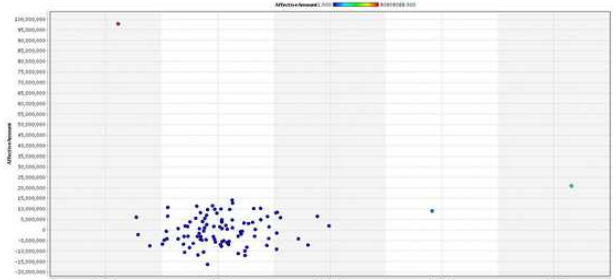

Figure 15. DBSCAN clustering for AffectiveAmount for abnormal Transactions

## IX. FDS & DSS Outputs

Developing FDS and DSS as software in order to help and advice inspector to inspect transaction faster with more accuracy is the last part of this research. When software load RapidMiner KDA model, model and its objects load in the memory and can interact directly with software and database as well. FDS has developed with Viusal Vb.Net 2010 and the Database Engine is Microsoft Sqlserver 2008.

In the DSS , two options provided in the software, first process the offline transactions, it mean we can run the system and check previous customer transactions by clicking on "Process Historical Data" button and second option process new transaction. For simulation purpose, we can add new transaction manually and process it. Definitely, in the online mode, the database updated automatically so no need to use this option. Figures 16-21show some FDS software outputs.

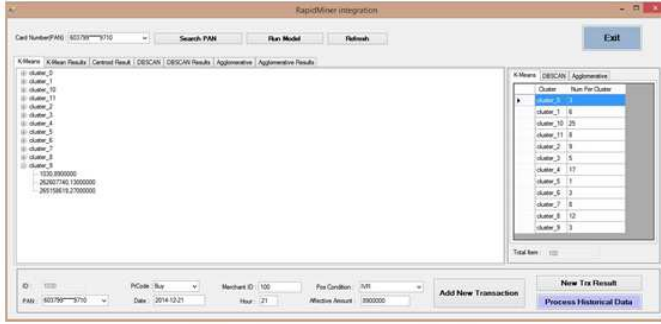

Figure 16- K-MEANS Tree Result

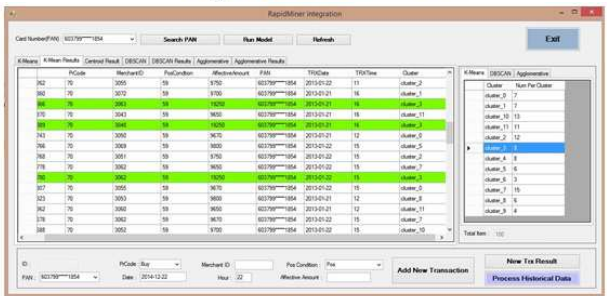

Figure 17 - highlighting cluster members

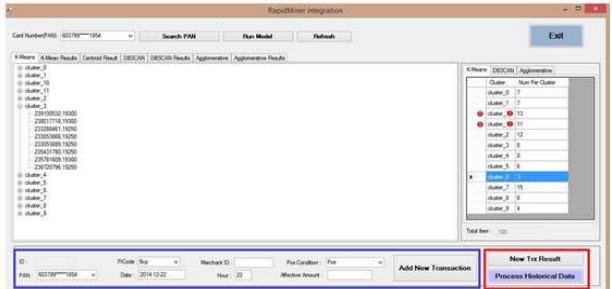

Figure 18- DSS options

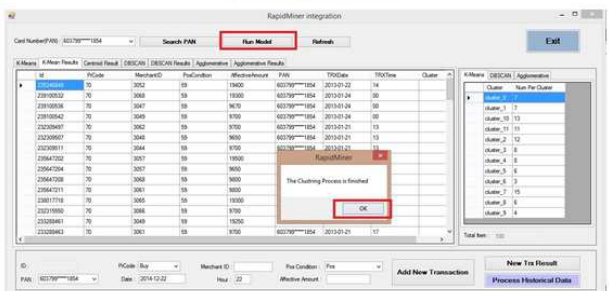

Figure 19- Applying KDA Model

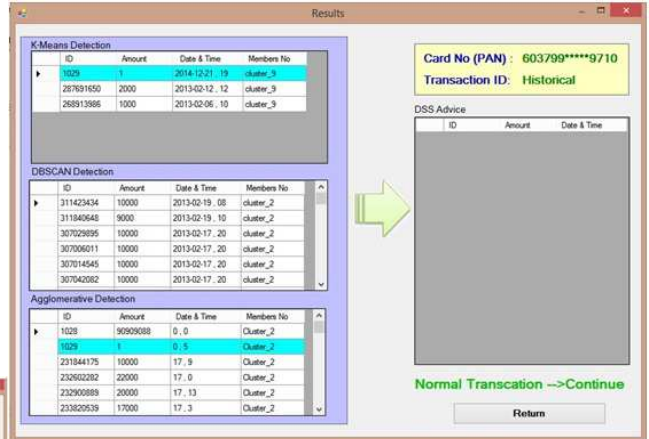

Figure 20- DSS result for normal historical dataset

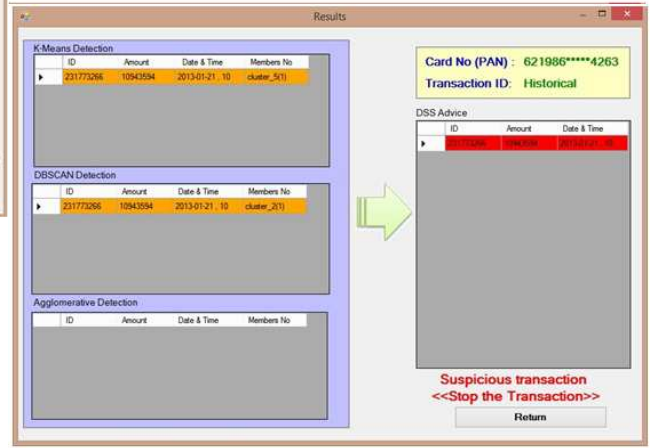

Figure 21- DSS result for abnormal historical dataset

## X. Conclusion

KDA model could improve consuming time processing and make three customer modeling in the same time to help detection suspious transaction in customer side better. Devloped FDS and DSS softwares can highlight and then classify the transaction with result of modeling.

The accuracy obtained by KDA modeling is 68.75% for dynamic online modeling and 81.25 % for historical or offline modeling and seemed it is competitive with other algorithms in this area.